# Single-shot and single-spot measurement of laser ablation threshold for carbon nanotubes

Vasily N. Lednev[1]*, Sergey M. Pershin[1], Elena D. Obraztsova[2], Sergey I. Kudryashov[3], Alexey F. Bunkin[1]

[1]Wave Research Center, Prokhorov General Physics Institute, Russian Academy of Sciences, Moscow, Russia

[2]Natural Sciences Center, Prokhorov General Physics Institute, Russian Academy of Sciences, Moscow, Russia

[3]Lebedev Physics Institute, Russian Academy of Sciences, Moscow, Russia

*Corresponding author: lednev@kapella.gpi.ru

**Abstract**

A simple and convenient procedure for single-shot, single-spot ablation threshold measurement has been developed. It is based on the employment of cylindrical lens to obtain elliptical Gaussian laser spot. The ablated spot chords which are parallel to the minor axis were measured across the spot major axis which is proportional to the fluence cross-section thus providing wide range dependence of damaged spot size versus fluence in one spot measurement. For both conventional and new-developed procedures the ablation threshold for typical Nd:YAG laser parameters (1064 nm, 10 ns) has been measured as $50\pm5$ mJ/cm$^2$ which is one order of magnitude lower than that for a bulk graphite.

A threshold for a laser-induced ablative material removal is a key governing parameter for a number of applications, such as a laser sampling for chemical analysis, a pulsed laser deposition of thin films, a new material synthesis, a coatings patterning, a laser machining and a laser systems development. Currently, several methods are used for evaluation of such laser ablation threshold: the ablative damage spot measurements [**1-3**] a detection of drastic ablative plasma emission appearance [**4**], a Langmuir probe and a mass-spectrometry detection of charged species yield [**5**].

Carbon materials, such as pyrolytic graphite, carbon nanotubes, and graphene are of key importance for both nanoscience and nanotechnology [6,7]. Laser ablation of graphite offers a simple way to produce many of the carbon allotropes while a post-production laser interaction with these materials provides their new functionalities. Particularly, bundles of single-wall carbon nanotubes (SWCNTs) is a promising material for the efficient field emitters [8], especially, in case when SWCNT thin films with their low ablation threshold are patterned by laser ablation [9-11] SWCNT thin films were also used as the passive light modulators for femtosecond lasers [12,13]. A laser lithography spatial resolution can be improved when SWCNTs are placed on the sample surface as the seeding "ablation centers" which initiate the substrate etching [14]. CNTs were proposed as the specifically adhesive, highly absorbing species for nanosurgery due to its ability to trigger ablation in the nanoscale intracellular regions [15]. In many such applications, laser ablation threshold provides a measure of optical and thermal SWCNT strength, being crucial both for their non-destructive exploitation and ablative processing.

This paper presents a novel simple procedure for the single-shot, single-spot measurements of laser ablation threshold for SWCNT and well-characterized pyrolytic graphite as a reference material. Conventional procedure for the threshold measurements assumes, as a pre-requisite, a Gaussian laser beam focused with a spherical lens and several laser shots at different pulse energies yielding in variable circular spots for the threshold measurements [1]. The peak-on-axis fluence at which the ablation is initiated is related to the ablation spot diameter due to the Gaussian beam profile. The squared spot diameter is then proportional to the natural logarithm of the peak-on-axis fluence. By measuring the spot diameter, one can derive the ablation onset fluence.

Meanwhile, potentially a two-dimensional fluence distribution provided by a focusing cylindrical lens enables to perform a single-spot measurement of the ablation thresholds through analysis of two uneven Gaussian beam profiles. By measuring ablated spots chords which are parallel to minor spot axis across the major axis and comparing with known fluence distribution for the major axis the ablation threshold can be calculated. The proposed technique is a 2D analog of conventional procedure with a continuous variation of fluence. The suggested method is similar to the microline technique, which has been developed previously for a fast elemental mapping by the laser induced breakdown spectroscopy [16]. Single-wall carbon nanotubes have been chosen as a test sample for the technique verification since this is a prospective material for the different applications discussed above. The standard wide spread solid state pulsed lasers can be effectively used for the laser ablation of such material for industry applications but the laser ablation threshold values for nanosecond pulses is absent in the literature.

Our experimental setup was described in details previously [17]. A Nd:YAG (1064 nm, 10 ns, Gaussian beam, 9 mJ/pulse) laser beam was focused onto a SWCNT sample by either a spherical lens ($F = 120$ mm), or a cylindrical lens ($F = 110$ mm) along the normal to the surface. All ablation tests were performed on a fresh sample surface as the single-shot exposures, using a manual translation stage to shift the sample between the succeeding shots. The laser pulse energy was monitored using a thermopile power meter (Oriel 70260) and attenuated by the neutral density filters. A laser beam profile at the sample surface was acquired by means of a CCD camera (PointGrey Dragonfly2). Bright-field images of the ablated spots were taken using an optical microscope equipped with a CCD camera.

The single-wall carbon nanotubes were synthesized in He atmosphere by arc discharge between two graphite electrodes with Ni catalyst [18]. The SWCNT content in the raw material was about 20%., but was then enriched up to 80% through ultrasonic treatment and ultracentrifugation [19]. Then, a SWCNT film was deposited on a filter, dried and characterized. The sample prepared was a semiconductor SWCNT with an average diameter of the synthesized tubes about 1.4 nm, as shown by high-resolution transmission electron microscopy and Raman spectroscopy [18].

In our experiments, laser ablation was defined as any permanent change of optical reflectance of a sample surface and several procedures were used to measure its threshold. In the first procedure, a well-characterized Gaussian focal spot was used to relate the observed damage spot radius to the corresponding threshold fluence [1], requiring, however, high accuracy when adjusting a beam profiler in the sample surface position.

The second way is to measure the peak-on-axis fluence that initiates the ablation onset [3]. Supposing that the laser beam is described by a Gaussian spatial and temporal profile:

$$I(r,t) = I_0 \exp(-2r^2/w^2)\exp(-t^2/\tau^2), \qquad (1)$$

where $I_0$ is the peak-on-axis irradiance (the power density), $r$ is the radial coordinate, w and $\tau$ are the spatial and temporal radii at the $1/e$- level of irradiance, respectively, $t$ is the time variable. Then, the spatial distribution of the fluence (energy density) is given by

$$F(r) = \int_{-\infty}^{+\infty} I(r,t)dt = F_0 \exp(-2r^2/w^2), \qquad (2)$$

where $F_0 = \sqrt{\pi}\tau I_0$ is the peak-on-axis fluence. The energy of the laser pulse can be determined by the fluence integration over the coordinates: $E = 2F_0/\pi w^2$. If the laser fluence exceeds the ablation threshold $F_{th}$ then the damaged spot will be formed with radius $r_{th}$: $F_{th} = F_0 \exp(-2r_{th}^2/w^2)$. Plotting the squared radii of the damages spots versus logarithm of the peak-on-axis fluence (or pulse energy, see figure 1) leads to a linear dependence:

$$r^2 = \frac{w^2}{2}\ln\left(\frac{F_0}{F(r)}\right). \qquad (3)$$

The peak-on-axis fluence, initiating ablation, is then derived as its abscissa intersect. This conventional method assumes a use of a normally incident axially symmetrical Gaussian beam focused by a spherical lens.

Meanwhile, additional information on ablation threshold can be extracted from the single-spot measurement, when two-dimensional fluence distribution will be used for laser ablation. For example, a cylindrical lens focusing will result in a two dimensional fluence profile with the different vertical cross-sections described by Gaussian widths ($w_x$ and $w_y$):

$$F(x, y) = F_0 \exp(-2x^2 / w_x^2) \exp(-2y^2 / w_y^2). \quad (4)$$

In such a case, the damaged spot has an elliptical contour with a large difference between the major and minor axis lengths (figure 2). As a result, its major axis $x$, which corresponds to the Gaussian beam cross-section with the larger width, can be employed as a continuous, low-gradient peak-on-axis fluence distribution. Then, the other, minor axis $y$, can be readily used for the threshold detection. Particularly, the squared lengths of the damage spot chords, that are parallel to the minor axis, can be expressed as:

$$y^2 = \frac{w_y^2}{2} \ln\left( F_0 \exp\left(\frac{-2x^2}{w_x^2}\right) / F(x,y) \right) = \frac{w_y^2}{2} \ln\left( \frac{F(x)}{F(x,y)} \right). \quad (5)$$

The term $F(x)$ is a laser beam fluence distribution which is known from the preceding beam profile measurements thus the ablation threshold $F_{th}$ is detected as abscissa intersect of $y^2$ to the zero (figure 3). The laser ablation threshold for the SWCNT sample was determined by all three procedures described above, with the results for the first two procedures presented in figure 1. For the first procedure, a comparison of the damaged spot dimensions with the horizontal cross-section of the fluence profile (figure 1a,b) gives the SWCNT ablation threshold of 60 mJ/cm$^2$ with an accuracy of 30%.

The second procedure employing $r^2 \sim \ln(E)$ dependence (see Eq.3) for the different ablative damage spot radii r and laser pulse energies E (figure 1c) provides the ablation thresholds for the SWCNT and highly oriented pyrolytic graphite (HOPG) samples, where the HOPG one was used as a well-known reference sample. The measured thresholds for SWCNT and graphite were 50 and 360 mJ/cm$^2$, respectively, with the latter threshold in a good agreement with the previously reported values [**20-22**]. However in the study by Chae et al. the SWCNT were not damaged by nanosecond pulse with fluence 590 mJ/cm$^2$ that is one order larger compared to our results. This difference can be explained by differences in SWCNT types (semiconducting vs. metallic type) and bulk sample forms (the ravel of nanotubes vs. nanotubes deposited on the quartz). If sample was a metallic SWCNT than a much higher ablation threshold should be obtained because of stronger adherence to the substrate.

A new proposed technique uses the cylindrical lens for the threshold measurements both in the focal plane and 20 mm upstream (figure 2) in order to compare the accuracy of the single spot measurements for different spot ellipticity. The magnified view of the second ablation spot in figure 2c and the corresponding cross-section of the beam profile in figure 2d are presented in figure 3(a,b). Here, the ablation damaged spot chords radii ($r_i=D_i/2$ coordinate $y_i$) were plotted as a function of laser fluence both for left and right parts of the spot. The resulting curve in figure 3c slightly deviates from the linear function near the SWCNT ablation threshold of 51 mJ/cm$^2$, which is equal to the value obtained using the conventional procedures (figure 1).

Surprisingly, the measured SWCNT ablation threshold was seven times lower than that for HOPG. While heat conduction characteristics of individual nanotubes is several times superior than that for graphite [23], the enhanced ablation of SWCNT may be related to significantly lower effective bulk thermal conductivity of typical SWCNT pellets due to their chaotic nanotubes packing with a few point contacts per each tube [24]. The energy transfer between SWCNTs in such bulk sample is extremely low resulting in much higher temperature of SWCNT and its destruction and plasma formation. Additionally, the SWCNT heat capacity is significantly lower compared to that for graphite [25] resulting in much higher temperature during first moments of laser ablation at similar deposited energies per atom compared to graphite.

This work was partially supported by the Russian Foundation for Basic Research (RFBR) under projects 10-02-00792, 10-08-00941, 11-02-00034, 11-02-01202 and 12-02-31398.

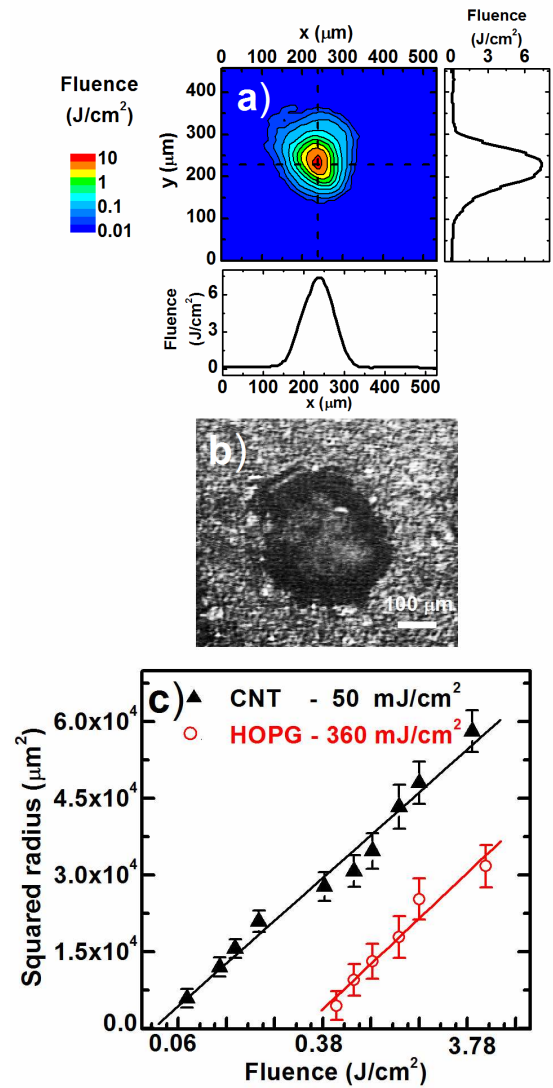

**Figure 1.** Laser ablation threshold measurements for SWCNT and HOPG by the conventional procedure: a) laser beam profile at the focal plane acquired by a CCD camera; b) optical image of the ablative crater on the SWCNT sample; c) $r^2 \sim \ln(F)$ curve (see details in the text).

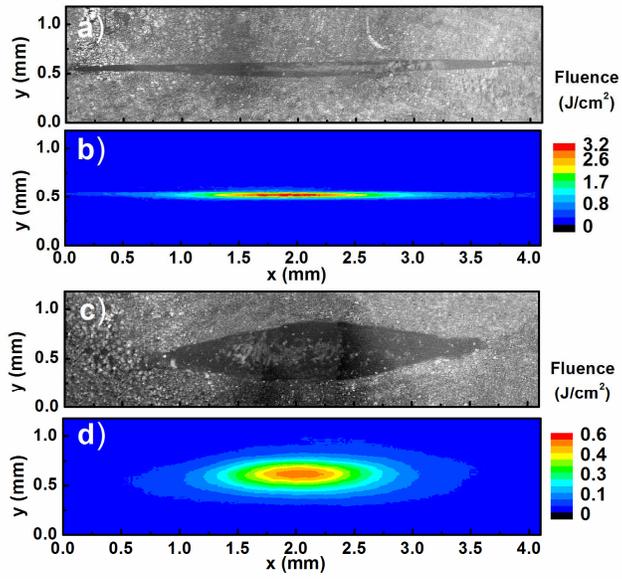

**Figure 2.** Images of ablation spot and beam profiles in the focal plane (a, b) and 20 mm upstream (c, d).

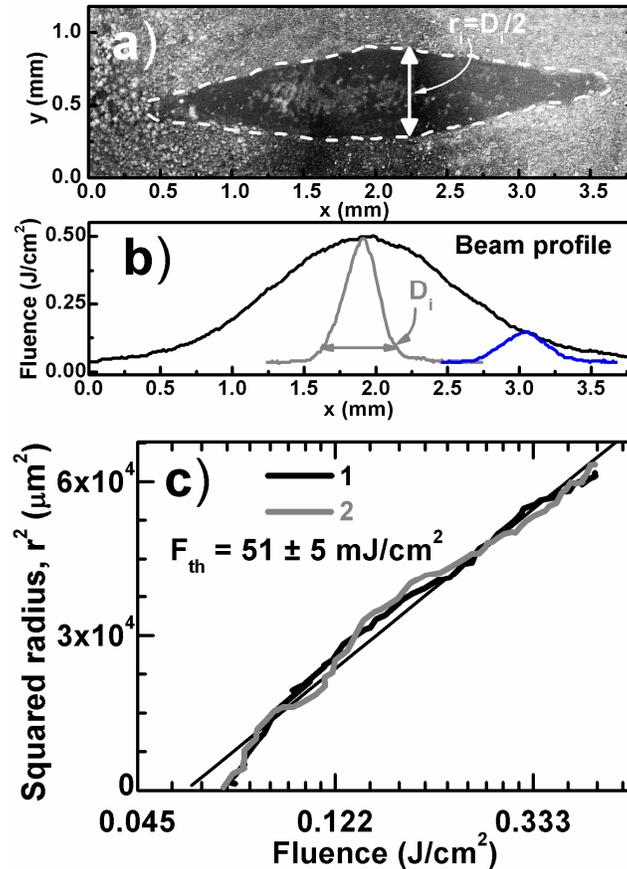

**Figure 3.** Laser ablation threshold measurement by a single spot single shot technique: a) image of the ablative damage spot and corresponding spot contour (dashed line) for the damage spot chords $D_i$ measurements; b) laser beam profile cross-sections from figure 2c: the cross-section corresponding to the major spot axis representing variable peak-on-axis fluence (black), the cross-sections corresponding to the minor spot axis for the damage chords measurements (grey for center, blue for wing); c) single spot ablation threshold measurements via linear fitting of the dependence of the squared damage chords ($r_i$) on fluence. The curves 1 (black) and 2 (grey) correspond to the experimental measurements over the right and left sides of the damaged spot.